\documentstyle[12pt]{article}
\Huge
\normalbaselines
\catcode `\@=11
\@addtoreset{equation}{section}

\def\section{\@startsection {section}{1}{\z@}{-3.5ex plus -1ex minus
     -.2ex}{2.3ex plus .2ex}{\normalsize\bf}}
\def\subsection{\@startsection{subsection}{2}{\z@}{-3.25ex plus -1ex minus
 -.2ex}{1.5ex plus .2ex}{\normalsize\bf}}

\def\@cite#1#2{${}^{\mbox{\scriptsize#1\if@tempswa , #2\fi}}$}
\def\thebibliography#1{\section*{References\markboth
  {REFERENCES}{REFERENCES}}\list
  {\arabic{enumi}.}{\settowidth\labelwidth{[#1]}\leftmargin\labelwidth
  \advance\leftmargin\labelsep
  \usecounter{enumi}}
  \def\newblock{\hskip .11em plus .33em minus -.07em}
  \sloppy
  \sfcode`\.=1000\relax}
 
\catcode `\@=12

\begin{document}

\vspace*{2.cm}
\noindent

{\bf The coherent states: old geometrical methods in new quantum clothes   }
\vspace{1.3cm}\\
\noindent \hspace*{1in} \begin{minipage}{13cm}
 Stefan Berceanu $^{1,2}$ \vspace{0.3cm}\\
 $^{1}$  Equipe de Physique Math\'ematique et
G\'eom\' etrie, Institut de Math\' ematique de Paris Jussieu CNRS,
Universit\' e Paris
7-
Denis Diderot, case 7012,
 tour 45-55,
5-\`eme \'etage, 2 place Jussieu, 75251- Paris Cedex 05, France, E-mail:
Berceanu@Mathp7.Jusssieu.Fr\\
$^{2}$ Permanent address: Institute of Atomic
Physics, Institute of Physics and Nuclear Engineering,
Department of Theoretical Physics, P. O. Box MG-6, Bucharest-Magurele,
Romania;
 E-mail: Berceanu@Roifa.Bitnet \\
\makebox[3mm]{ }\\
\end{minipage}

\vspace*{0.3cm}

\noindent
   {\it Abstract }  A geometric characterization of transition amplitudes
   between coherent
states, or equivalently,
of the hermitian scalar product of holomorphic
cross sections in the associated $D_{\widetilde {\bf M}}$ - module,
in terms of the embedding  of the coherent state manifold $\widetilde {\bf M}$
into a projective  Hilbert space is proposed. Coherent state manifolds endowed
with a homogeneous kaehler structure are considered. Using the coherent state
approach, an effective method to find the cut loci  on symmetric manifolds and
generalized
symmetric manifolds $\widetilde {\bf M}$
is proposed. The CW- complex structure of coherent state manifolds of flag type
is discussed. Recent results of Anandan and Aharonov
are commented vis-\` a-vis of last century constructions in projective
geometry.
Calculations with significance in the coherent state approach furnishe explicit
proofs of the results announced by Y. C.Wong
on conjugate locus in complex Grassmann manifold.

\vspace{.25in}


Let a quantum system with symmetry, i.e.
a triplet
$({\bf K},G,\pi)$, where  $\pi$ is an unitary irreducible
representation of the group $G$ on the Hilbert space ${\bf
K}$. Let the orbit $\widetilde {\bf M} ={\widetilde
\pi (G)}\vert \widetilde {\psi_0>}~$ , where $\vert  \psi_0> \in \bf
K, \xi \vert  \psi_0>=\vert \widetilde\psi_0>~\in \bf {PK}$. Then
$\widetilde {\bf M} \approx G/K $ (diffeomorphism), where $K$ is
the stationary group of the state $\vert \widetilde\psi_0>$. If
$i:\widetilde {\bf M} \hookrightarrow {\bf PK}$ is an embedding ,
then $\widetilde {\bf M}$ is called {\it coherent state
manifold}. If $\vert  \psi_0> \equiv \vert j>$ is an extremal weight
vector and G is compact connected simply connected, then
 $\widetilde {\bf M}$  is K\"ahler and $\pi_j$ is given by the Borel-
Weil-Bott theorem.If a local section $\sigma:\widetilde {\bf
M}\rightarrow{\it S(\bf K)}$ in the unity sphere in $\bf K$ is constructed,
then the set ${\bf M} =\sigma(\widetilde {\bf M})$ is named {\it
coherent vector manifold}. $\bf M$ is the holomorphic line
bundle associated by a holomorphic character $\chi$ of the
parabolic subgroup P of the complexification $G^c$ of G.

The coherent states are introduced as

$$\vert
Z,j>=\exp\sum_{{\varphi}\in\Delta^+_n}(Z_{\varphi}F^+_{\varphi})\vert
j> ,\widetilde{\vert
Z>}=< Z \vert Z>^{-1/2}\vert Z> ,\eqno (*)$$
where $\Delta^+_n$ are the positive non-compact roots. If $ Z=(Z_ \varphi )
\in {\bf C}^d$, where d is the dimension of $\widetilde {\bf M}$, then Z are
local
coordinates in the neighborhood $\it V_0 \subset \widetilde {\bf M} $ around
$Z=0$. Here $F^+_{\varphi}\vert j>\neq  0,F^-_{\varphi}\vert j> = 0,~  \varphi
\in\Delta^+_n. $
Then $<\widetilde { Z^\prime} \vert\widetilde { Z}>$ is :i) the hermitian
scalar
product of holomorphic sections in the line bundle ${\bf M}$ associated by
$\chi$ to the
principal holomorphic bundle $P\rightarrow G^c\rightarrow G^c/P$, or ii) the
hermitian scalar product of sections with base $\widetilde {\bf M}$ in the
$D_{\widetilde {\bf M}}$ - module of differentiable operators on $\widetilde
{\bf M}$. Here $G^c/P$ is a flag manifold.

{\bf Problem 1:} find a geometric meaning of the transition probability on
coherent state manifold.

{\bf Proposition 1:} Let $\widetilde{\vert
Z>}$ as in $(*)$, where Z parametrize the coherent state manifold
in the ${\it V}_0 \subset \widetilde {\bf M} $ and let the embedding $i:
\widetilde {\bf M}
\hookrightarrow {\bf PK}$. Then the angle

$$\theta \equiv arccos \vert <\widetilde { Z^\prime} \vert\widetilde { Z}>\vert
,$$
is equal to the Cayley distance on the geodesic joining $i(Z'),i(Z),$ where $
Z', Z \in {\it V}_0$,
$$\theta = d_c(i(Z'),i(Z)).$$

More generally, it is true the following relation (Cauchy formula)

$$<\widetilde { Z^\prime} \vert\widetilde { Z}>=(i(Z'),i(Z)).$$

{\it Proof:} The holomorphic line bundle {\bf M} of coherent vectors is the
pull
back $i^*$ of
the hyperplan bundle of {\bf {PK}}, the dual bundle of the tautological
line bundle of ${\bf PK^*}$.

Here (.,.) is the scalar product in {\bf K}. If $\xi :{\bf K}\backslash  \{0\}
\rightarrow
{\bf PK},\xi : \omega \rightarrow [\omega ]$, then the Cayley (1859) distance
is

$$ d_c([\omega'],[\omega ])=arccos{ \vert (\omega ',\omega )\vert \over  \|
\omega '\| \| \omega \| }.$$

For the $\infty$-dimensional case, see Kobayashi (1959), which also gives
conditions for
the existence of the embedding $i$. See also Rawnsley (1977).

{\it Comment:} The  Cayley distance has been used in Quantum Mechanics by Wick
(
1967), Fivel
(1973) and recently by Anandan and Aharonov  (1990). The Cayley distance is
useful in the
geodesic approach. The (Bargmann) distance $d_b$, used by Prevost and Vall\' ee
(1980)
in the context of  coherent states,

$$d^2_b([\omega'],[\omega ])=2(1-cosd_c([\omega'],[\omega ])),$$
is equivalent with $d_c :{2\sqrt {2}}/ \pi \leq d_b\leq d_c.$

{\bf Problem 2:} find those manifolds $\widetilde {\bf M} $ for which the angle
$\theta \equiv arccos \vert <\widetilde { Z^\prime} \vert\widetilde { Z}>\vert
$
 is a distance on $\widetilde {\bf M} $.

{\bf Proposition 2:} Let $\widetilde {\bf M} $ be a coherent state manifold
parametrized as in
$(*)$. Then the angle $\theta \equiv arccos \vert <\widetilde { Z^\prime} \vert
\widetilde { Z}>\vert $
is a distance on $\widetilde {\bf M} $ iff $\widetilde {\bf M} $ is a symmetric
space of rank 1.

{\it Proof:} The problem is reduced to that of two-point homogeneous
spaces, which are known ( Wolf).

{\it Comment:} Generally, the distance $\delta \geq \theta $, but
infinitesimally, $d\delta
=d\theta .$ Let us illustrate this on the complex Grassmann manifold $ G_n(\bf
C
^{n+m})$. Then
$$cos\theta =cos\theta _1...cos\theta _n, \theta =d_c(i(0),i(z)),
i : G_n(\bf C^{n+m})\hookrightarrow
{\bf CP}^{\scriptstyle {\left(\!\begin{array}{c}m+n\\n\end{array}\!\right) -1}}
,$$
{\it i} is the
Plucker embedding,
$\theta _i, i=1,...,n$ are the stationary angles of the two n-planes in ${\bf
C}
^{n+m}$
(Jordan (1875)),
$\delta ^2= \sum_{i=1}^{n}\theta _i^2 , (\theta _i <\pi /2,
i=1,...,n)$ (Rosenfel'd (1941)).

{\bf Problem 3:} find a geometric meaning of Calabi's diastasis (1953), used by
Cahen, Gutt,
Rawnsley
(1993) in the context of coherent states ,
$D(Z',Z)=-2ln \vert <\widetilde { Z^\prime} \vert\widetilde { Z}>\vert .$

{\bf Proposition 3:} The diastasis distance $D(Z',Z)$ between $Z', Z \in {\it
V}
_0
 \subset \widetilde {\bf M}$ is related to the geodesic distance $\theta =
d_c(i
(Z'),i(Z))$,
 where $i:\widetilde {\bf M} \hookrightarrow {\bf PK},$ by

 $$D(Z',Z)= ln cos^{-2}\theta .$$

 If $\widetilde {\bf M}_n$ is noncompact and $i':\widetilde {\bf M}_n
 \hookrightarrow {\bf PK}^{N-1,1}=
  SU(N,1)/S(U(N)\times U(1)) (i:\widetilde {\bf M}_n \hookrightarrow {\bf PK})$
and $\delta _n (\theta
  _n)$ is the length of the geodesic joining $i'(Z'),i'(Z) (resp. i(Z'),i(Z))$,
then $$cos\theta _n=
  cosh^{-1}\delta _n=e^{-D/2}.$$

  {\it Proof:} cf. Prop. 1.

{\bf Problem 4:} characterize the relationship of the number $N$ and the
manifold $\widetilde {\bf M}$
 in the embedding $i:\widetilde {\bf M} \hookrightarrow {\bf CP}^{N-1}.$

 {\bf Proposition 4:} For coherent state manifolds $\widetilde {\bf M} \approx
 G/K $ which have a flag
 manifold structure, the following 7 numbers are equal:

 1) the maximal number of orthogonal coherent vectors on $\widetilde {\bf M}$;

 2) the number of holomorphic global sections in the holomorphic line
bundle{\bf
 M} with base
 $\widetilde {\bf M}$;

 3) the dimension of the representation in the Borel-Weil-Bott theorem;

 4) the minimal $N$ appearing in the Kodaira embedding theorem,
 $i:\widetilde {\bf M} \hookrightarrow {\bf CP}^{N-1}$;

 5) the number of critical points of the energy function $f_H$ attached to a
 Hamiltonian $H$
 linear in the generators of the Cartan algebra of G, with inequal
coefficients;

 6) the Euler- Poincar\' e characteristic of $\widetilde {\bf M} \approx G/K ,
\
chi
 (\widetilde {\bf M})=
 [W_G]/[W_H]$, where $[W_G]= card W_G$,and $ W_G$ is the Weyl group of G;

 7) the number of Borel-Morse cells which appear in the CW- complex
 decomposition of $\widetilde {\bf M}$.

 {\it Proof:} Use theorems 1, 2 in S.B. and A.C.Gheorghe (1987),
 where it is proved that$f_H$
 is a perfect Morse function, and also the Cauchy formula. Remark that
 $\chi (G/K) >0$ iff $Rank G= Rank K$( cf. Hopf, Samelson (1941)).

 {\it Comment :} The Weil prequantization condition is the condition to have a
 Kodaira embedding,
 i.e. the algebraic manifold to be Hodge.

 {\bf Problem 5:} find a relationship between geodesics and coherent states.

 {\bf Proposition 5:} For a d-dimensional manifold $ G/K$ equipped with
 hermitian symmetric
 space structure, the parameters $B_\varphi $ in formulas of coherent vectors
 $$\vert
B,j>=\exp\sum_{{\varphi}\in\Delta^+_n}(B_{\varphi}F^+_{\varphi}-{\bar
B}_{\varphi}F^-_{\varphi})\vert j> ,
\eqno (**)$$
$$\vert
B,j>\equiv \widetilde  {\vert Z,j>}=< Z,j \vert
Z,j>^{-1/2}\exp\sum_{{\varphi}\i
n\Delta^+_n}(Z_{\varphi}F^+_{\varphi})\vert
j> ,\eqno (***)$$
are normal coordinates in the normal neighborhood ${\it V}_0\approx {\bf C}^d$
around $Z=0$.
The conjugate locus of the point  $Z=0$ is obtained annulating the Jacobian of
t
he
transformation $Z=Z(B)$.

{\it Proof:} See S.B. and  L.Boutet de Monvel (1993).

{\bf By-product:} The results announced by Wong (1968) on conjugate locus
in $ G_n(\bf C^{n+m})$ are proved.They were contested by T. Sakai in 1977. The
proof uses the Jordan angles.

{\bf Problem 6 :} characterize geometrically the polar divizor of $\vert 0>,$

$$\Sigma _0= \left\{\vert \psi > \vert \vert \psi >={\rm  coherent\mbox{~}
vecto
r},
<0\vert \psi >=0 \right\} .$$

Let $e$ be the unity element in  $ G/K, \lambda :G\rightarrow G/K$ the natural
 projection,
${\bf g=k\oplus m}$ the Lie algebra decomposition, $Exp: T_pM\rightarrow M$
the exponential mapping and $exp :{\bf g}\rightarrow G$. Let the condition
\newline

$ A1) \hspace{1.in} Exp\vert \lambda (e)=\lambda \circ exp\vert {\bf m} , $
\newline

i.e. the geodesics in $\widetilde {\bf M}$  are images of one-parameter
subgroups of G.
\newline
Thimm furnishes sufficient conditions for the manifold  $\widetilde {\bf M}$ to
verify
$A1)$, for example the  reductive manifolds, and in particular the
 symmetric spaces verifie  $A1)$ (cf. Cartan).

 In the next proposition, ${\bf CL}_0$ denotes the cut locus of $ 0\in
\widetilde
 {\bf M}$.

 {\bf Proposition 6:} Let $\widetilde {\bf M} \approx G/K $, and suppose the
 parametrization
$ (*)$ around $Z=0$ in $\it V_0$. Then $\widetilde {\bf M}=\it V_0\cup \Sigma_0
$ (disjoint union).
 If $A1)$ is true for the manifold $\widetilde {\bf M} $, then $\Sigma _0={\bf
 CL}_0.$

 Moreover, if $i$ is the embeeding $i:\widetilde {\bf M} \hookrightarrow
 {\bf PK}$, then

 $${\bf CL}_0=\left\{Z \in \widetilde {\bf M}\vert d_c(i(0),i(z))    =\pi /2
 \right\}.$$

 {\it Proof:} The theorems reffering to ${\bf CL}$ are from  Kobayashi, Nomizu
 Vol ll.
 See also S.Kobayashi in "Global differential geometry" (1989).
 \vspace{0.25in}

 This paper was presented at the Xl-th ICMP in Paris 1994. The author is
grateful to  Professor
 Anne Boutet de Monvel for interest and remarks. The constant
 supervision
 from Prof. L. Boutet de Monvel is kindly acknowledged. Disscusions  and
suggest
ions
 from Professors K. Teleman, C.A. Gheorghe, H. Scutaru, G. Ghika,
 B. Berceanu, M. Duflo, A. Kirilov, A. M. Perelomov, D. Simms and
 correspondence with Professors Th. Hangan, M. Berger and S. Kobayashi
are
 acknowledged. I also acknowledge the support from CNRS and IAP.
 \vspace{0.1in}

  References (selective list):
 \vspace{0.25in}

  A.Cayley, Phil. Trans. Royal. Soc. London 149 ({\bf 1859}) 61; C. Jordan,
 Bull. Soc. Math. France. t III ({\bf 1875}) 103; E.A.Rosenfel'd, Izv.Akad.Nauk
SSSR, ser.
 Mat. 5 ({\bf1941}) 353; E.Calabi, Ann.Math. 58 ({\bf 1953}) 1;S.Kobayashi,
Tran
s.Amer.Math.Soc.
 92 ({\bf 1959}) 267; G.K.Wick,  {\it On symmetry transformations}, in
"Preludes
 in
 Theoretical Physics", North Holand ({\bf 1967});J.Wolf, Spaces of constant
 curvature ({\bf
 1967});
 Y.C.Wong, Bull.Am.Math.Soc. 74 ({\bf 1968}) 240;  A.M.Perelomov, Commun.Math.
 Phys. 26 (
 {\bf 1972}) 222;
  D.I. Fivel, preprint Maryland 73-102 ({\bf 1973});J.Rawnsley, Quart.J.Math.
  Oxford 28 (
  {\bf 1977}) 403;
  T.Sakai, Hokkaido Math. J. 6
 ({\bf 1977})
  136;
  J.P.Prevost, G.Vall\' ee, Commun.Math.Phys.
 76 ({\bf 1980}) 289; A.Thimm, Ergod. Theory Dyn. Syst. 1 ({\bf 1981}) 495;
 S.Berceanu, A.Gheorghe, J.Math.Phys. 20 ({\bf 1987}) 2892;
  J.Anandan, Y.Aharonov, Phys.Rev.Lett.
 65 ({\bf 1990}) 1697 ; S.Berceanu, L.Boutet de Monvel,
 J.Math.Phys.34 ({\bf 1993}) 2353;  M.Cahen, S.Gutt, J.Rawnsley, Trans. Math.
 Soc. 337 ({\bf 1993 }) 73.
\end{document}